\documentclass[11pt]{article}

\usepackage[margin=1in]{geometry}
\usepackage{microtype}
\usepackage{setspace}      

\usepackage{graphicx}
\usepackage{booktabs}

\usepackage[round,sort&compress]{natbib}

\usepackage[hidelinks]{hyperref}
\usepackage[nameinlink,capitalise]{cleveref}

\usepackage{url}

\usepackage{tikz}
\usetikzlibrary{arrows.meta,positioning,shapes.multipart,backgrounds,fit,calc}

\AtBeginDocument{%
  \providecommand\BibTeX{{B\kern-.05em\textsc{i\kern-.025em b}\kern-.08em\TeX}}}

\newcommand{\keywords}[1]{\vspace{0.5em}\noindent\textbf{Keywords: }#1}
\providecommand{\Description}[2][]{} 

\newif\ifanon
\anonfalse

\title{Proactive AI Adoption can be Threatening: When Help Backfires}

\usepackage{authblk}

\ifanon
  \author{Anonymous Authors\\Affiliations withheld for review}
  \date{}
\else
  \author[1]{Dana Harari}
  \author[1]{Ofra Amir}
  \affil[1]{Faculty of Data and Decision Sciences, Technion - Israel Institute of Technology, Israel}
  \date{}
\fi

\begin{document}

\maketitle

\begin{abstract}
Artificial intelligence (AI) assistants are increasingly embedded in workplace tools, raising the question of how initiative-taking shapes adoption. Prior work highlights trust and expectation mismatches as barriers, but the underlying psychological mechanisms remain unclear. Drawing on self-affirmation and social exchange theories, we theorize that unsolicited help elicits self-threat, thereby reducing willingness to accept help, likelihood of future use, and performance expectancy of AI.  
We report two vignette-based experiments (Study~1: $N=761$; Study~2: $N=571$, preregistered). Study~1 compared anticipatory and reactive help provided by an AI vs. a human, while Study~2 distinguished between \emph{offering} (suggesting help) and \emph{providing} (acting automatically). In Study 1, AI reactive help was more threatening than reactive human help. Across both studies, anticipatory help increased user's self-threat and reduced adoption outcomes. Our findings identify self-threat as a mechanism through which anticipatory help, a proactive AI feature, may backfire, and suggest design implications to be tested in interactive systems. 
\end{abstract}



\section{Introduction}

Artificial intelligence (AI) assistants such as ChatGPT, GitHub Copilot, Microsoft 365 Copilot, and Anthropic are rapidly becoming part of everyday work practices. Field and experimental studies show that such tools can substantially increase productivity across domains. For example, research suggests that support from ChatGPT improved professional writing speed and quality \citep{noy_experimental_2023-1}, GitHub Copilot accelerated coding performance \citep{peng_impact_2023}, and generative AI increased customer support efficiency \citep{brynjolfsson2025generative} and consulting output quality \citep{dellacqua_navigating_2023}. Industry reports further support for these findings, highlighting how workplace AI assistants are being integrated at scale~\citep{butler_microsoft_2023}. 

These tools promise to boost productivity by assisting with a variety of tasks including drafting text, writing code, and generating insights. Their growing presence in workplace technologies means that employees increasingly encounter AI not as an optional tool but as an embedded collaborator. Yet realizing the benefits of these systems depends not only on technical capability but also on whether employees are willing to adopt and use them for their daily work. Importantly, Human Computer Interaction (HCI) research has established that employees are often reluctant to adopt new technologies, which can limit the performance benefits of AI at work. One potential avenue for increasing adoption is to design AI systems to be more proactive in when and how they provide help.

From a design standpoint, a central question concerns \emph{initiative}: should assistants wait for explicit help requests, or provide help proactively? For example, GitHub Copilot can generate code snippets in real time without being prompted, Gmail's Smart Compose fills in sentences as users type, and Microsoft 365 Copilot suggests edits or follow-up actions during meetings. Such proactive behaviors by AI are already widespread, yet their effects on adoption remain unclear. While prior work has documented that when the timing or level of AI initiative does not match users' preferences or context, trust and acceptance can suffer \citep{liao_what_2016,meurisch_exploring_2020}, the underlying \emph{psychological mechanisms} behind adoption decisions following proactive AI assistance are less well understood. This raises the question of why people reject proactive AI assistance even when it could be useful.  

To address this gap, we draw on organizational psychology and management research on helping. In managerial research regarding human–human collaboration, help can be \emph{reactive} (requested by the recipient) or \emph{anticipatory} (unsolicited, either suggested or delivered) \citep{nadler_role_1986,harari_when_2022}. While reactive help is consistent with reciprocity norms of social exchange theory \citep{cropanzano_social_2005,emerson_social_1976}, anticipatory help can threaten recipients' independence and competence and is more likely to threaten reciprocity norms. According to self-affirmation theory, this effect emerges because unsolicited help undermines people's positive self-view, producing defensive reactions \citep{steele_psychology_1988,sherman_psychology_2006}.  
In line with prior managerial research, we argue that a central mechanism behind employees' resistance to AI anticipatory help is self-threat~\citep{steele_psychology_1988}, when an employee experiences events that challenge their sense of themselves as good or successful. Threats to the self often involve a failure to meet social standards, which at work often revolves around meeting standards of competence and independence. These are most consequential for employees' positive sense of self at the workplace since the ability to do one's job well and independently is an important part of people's work identity. As such, in workplace settings, anticipatory help can challenge employees' positive sense of self and be threatening because it can convey that they are not able to complete their job without assistance, which will elicit feelings of inadequacy and be even more harmful when it is unexpected, such as the case of anticipatory help \citep{harari_when_2022}. As a result, proactive AI assistance may be threatening for employees' positive sense of self.

We theorize that these dynamics also apply to human–AI interaction: anticipatory AI help may be perceived as more threatening than reactive help, even when the actual assistance provided does not differ. Furthermore, in human--AI interaction it is more salient that anticipatory help can take different forms: \emph{offering} (the system suggests help but awaits user acceptance) and \emph{providing} (the system delivers help automatically). In what follows, we use \emph{proactive} as a broader term for AI systems that take initiative, and \emph{anticipatory help} to refer to the specific form of unsolicited help examined in our studies.

We investigate these dynamics through two vignette-based experiments with large samples. Across both studies, we examine whether \emph{self-threat} links AI initiative to user adoption outcomes. Study~1 used a 2 (help type: anticipatory vs.\ reactive) $\times$ 2 (help source: AI vs.\ human) design to test whether anticipatory help is more threatening than reactive help (H1), the mediation of self-threat for AI adoption outcomes (H2), and whether self-threat responses differ by help source (H3). Study~2, preregistered, extended this design by comparing reactive help to two forms of anticipatory help: offering and providing (H4). 

Results from Study 1 suggest that reactive AI help was rated as more threatening than reactive human help, suggesting that reciprocity norms buffer human but not AI helpers. Results from both studies show anticipatory help elicited significantly more self-threat than reactive help, which in turn reduced adoption outcomes (willingness to accept help, likelihood of future use, and performance expectancy of technology). In Study~2, both \emph{offering} and \emph{providing} anticipatory AI help increased self-threat relative to reactive help, while the two forms of anticipatory help (offering vs. providing) did not significantly differ. Together, these results provide support for self-threat as a mechanism through which anticipatory AI help may backfire by lowering adoption of AI at work.

This paper makes three contributions:
\begin{enumerate}
    \item We advance research on mixed-initiative interaction by developing a deeper understanding of employees' experiences with help from AI tools at work. Drawing on organizational theories, we explain when and why employees are more or less likely to adopt AI help at work, highlighting self-threat as a psychological mechanism linking AI initiative to adoption.
    
    \item We contribute to research on AI at work by identifying the distinct effects of anticipatory versus reactive AI help. We show that, although anticipatory help may increase employees' exposure to AI's performance benefits, it can also heighten feelings of self-threat relative to reactive help. These insights regarding AI help contribute to theory and practice regarding AI in organizations.
    
    \item We offer preliminary design insights on AI initiative by distinguishing between anticipatory help that offers help and anticipatory help that provides it automatically. This distinction extends current research on human-AI interactions by highlighting the importance of initiative in shaping users' psychological response, beyond accuracy and error costs.
\end{enumerate}

\section{Related Work}

Our work builds on three bodies of literature: HCI research on system initiative, organizational and psychological research on helping, and studies of AI adoption. Together, these bodies of work motivate our focus on \emph{self-threat} as a mechanism explaining when and why anticipatory AI help undermines user adoption.

\subsection{System Initiative in HCI}

HCI has long examined how initiative is negotiated between users and systems. Horvitz's classic work on mixed-initiative interfaces emphasized that system actions should be calibrated to confidence, user goals, and the cost of errors \citep{horvitz_principles_1999}. Amershi et al.'s guidelines \citep{amershi_guidelines_2019}  codified these insights into widely adopted principles for human–AI interaction, such as timing system actions appropriately and clarifying why the system acted. While these frameworks address \emph{when} initiative may be appropriate, they leave psychological consequences under-specified.

Prior studies on proactivity found that proactive features are often perceived as interruptive or intrusive.
\citet{liao_what_2016} conducted a field study of AI in the workplace and found that proactive assistance carries a risk of interruption and is often not utilized. A survey on expectations from proactive AI systems~\cite{meurisch_exploring_2020} found that users expect such systems to balance initiative with control, transparency, and context-awareness, and that personality traits and socio-demographic characteristics affect user expectations for proactivity. Reports from industry mirror these findings: proactive Copilot features can accelerate productivity but also overwhelm workers when assistance is unsolicited \citep{butler_microsoft_2023}. Recent findings have further emphasized the need to balance worker and system initiative in AI-supported work. For example, He et al. identify task dimensions—such as process and social consequence, task familiarity, and complexity—that shape when workers prefer to retain initiative and control versus delegate them to AI systems \citep{He2023Rebalancing}.
These studies demonstrate that initiative-taking is consequential, but they leave underexplored the psychological processes through which it affects user adoption.

\subsection{Barriers to AI Adoption}

Even when AI systems deliver measurable performance gains, adoption is not guaranteed. Large-scale field studies show productivity improvements across domains, including writing \citep{noy_experimental_2023-1}, customer support \citep{brynjolfsson2025generative}, programming \citep{peng_impact_2023}, and consulting \citep{dellacqua_navigating_2023}. Yet, employees often remain reluctant to rely on AI tools. Research highlights barriers such as algorithm aversion \citep{dietvorst_algorithm_2015}, trust and transparency concerns \citep{chiu_hearts_2021,glikson_trust_2020}, and social consequences of adoption \citep{tang_no_2023,man_tang_when_2022,berg_capturing_2023}. Explanations and transparency features can build trust \citep{ehsan_explaining_2021}, but they may not fully address how AI assistance is experienced in relation to professional norms and self-relevant concerns. Recent research shows that professionals in high-responsibility roles (e.g., judges and air traffic controllers) raise substantial concerns about how AI assistance may affect reliability, trust, skill requirements, and responsibility in their work, rather than treating AI as a neutral productivity aid \citep{Solovey2025InteractingAI,Schon2025ClearedTakeoff}. This suggests that adoption decisions are influenced not only by technical performance, but also by how AI help aligns with professional norms and expectations.

\subsection{Psychological Perspectives on Help-Seeking}

Psychological research on helping offers perspective on why AI proactivity may backfire. Previous research distinguishes between \emph{reactive help} (requested by the recipient) and \emph{anticipatory help} (unsolicited: offered or provided without an explicit request) \citep{nadler_role_1986,nadler_help-seeking_1991}. Although reactive help is often welcomed, anticipatory help can feel intrusive and threatening. ~\citet{harari_when_2022} show that anticipatory help, especially from higher-status helpers, increases recipient self-threat and reduces evaluations of the helper and willingness to accept help. Relatedly, ~\citet{lee_benefits_2019} show that anticipatory help is less likely to relate to gratitude and positive work outcomes.

In this paper, we argue that anticipatory help may be more likely than reactive help to elicit \emph{self-threat} because it arrives without being requested and can therefore be construed as implying reduced competence, dependence on others, or diminished standing on an identity-relevant task. Importantly, help can be instrumentally beneficial while still carrying psychological costs. Recipients may value the help itself while reacting negatively to what accepting that help seems to imply about them at work.

Two theoretical perspectives help explain why anticipatory help may elicit self-threat. \emph{Social exchange theory} suggests that solicited help is more likely to align with reciprocity norms, whereas unsolicited help can violate those norms, leading recipients to devalue the helper \citep{cropanzano_social_2005,emerson_social_1976}. \emph{Self-affirmation theory} highlights a complementary mechanism: unsolicited help can be threatening because it may signal that the helper does not view the recipient as competent or independent enough to complete the task on their own~\citep{steele_psychology_1988,sherman_psychology_2006}. Such threat can trigger defensive reactions, including rejecting help or lowering evaluations of the helper. Related work on feedback interventions similarly shows that unsolicited or negative feedback often undermines motivation and other positive outcomes \citep{ilgen_bearing_2000,kluger_effects_1996}. These dynamics are especially likely when the task is highly identity-relevant, since help in such contexts is more likely to be seen as a reflection of the self \citep{harari_when_2022}.

\subsection{Extending to Human--AI Interaction}

We theorize that these dynamics extend to human--AI interaction. Anticipatory AI help, whether suggesting help (\emph{offering}) or taking action automatically (\emph{providing}), may challenge employees' positive sense of self. It may do so by implying reduced competence, independence, or autonomy, thereby eliciting self-threat and reducing willingness to adopt AI tools. By contrast, reactive AI help is more consistent with user control, since employees decide why, when, and how to seek help. As a result, it will be less likely to trigger self-threat.

Importantly, AI helpers may differ from human helpers. Reciprocity norms that often buffer human-to-human help may not apply to AI systems, meaning that even reactive AI help could feel more threatening than reactive human help. This builds on the Computers Are Social Actors (CASA) paradigm, which shows that people apply social rules such as reciprocity to computers, but often in domain-dependent or different ways \citep{nass_computers_1994,nass_machines_2000}. Recent HCI studies suggest that human--AI interactions draw on familiar social processes while also introducing distinct ones: for example, trust formation can resemble human--human dynamics but diverges due to opacity and lack of reciprocity \citep{glikson_trust_2020}, and mismatches between user expectations and system initiative can undermine trust in unique ways \citep{meurisch_exploring_2020}. 

In addition, research on HCI in the workplace highlights that interactional factors beyond initiative shape how AI help is experienced. For example, practitioners expressed strong resistance to AI systems that issue directive recommendations, while preferring analytical support that does not prescribe action \citep{Ma2025DontTellMe}. Relatedly, Rinott and Shaer emphasize that temporal characteristics such as interaction pace, continuity, and rhythm play a central role in shaping human--AI collaboration at work, influencing how AI contributions are seen over time \citep{Rinott2024TemporalAI}.

These findings also suggest that resistance to proactive AI may arise through multiple, partially overlapping psychological pathways. In particular, anticipatory AI help can alter users' sense of autonomy, control, and agency by changing who initiates action, who directs the workflow, and how much discretion users retain over task execution. In this paper, we do not treat these constructs as competing explanations to be ruled out. Rather, we focus on \emph{self-threat} as a specific mechanism that may operate alongside them. As such, we argue that anticipatory help will be threatening, especially in identity-relevant tasks, due to challenging employees' sense of competence or independence.

Moreover, emerging work that has relied on managerial research about helping has begun to explore how proactive vs. reactive AI behavior shapes perceptions of competence and system satisfaction. In particular, recent Information Systems research highlights similar dynamics. For example, Diebel et al.~\citep{diebel_when_2024} show that proactive AI support can reduce users' competence-based self-esteem, particularly when the AI demonstrates high levels of knowledge. Their findings reinforce our argument that unsolicited initiative can undermine psychological needs for competence and autonomy, even when the assistance is accurate or useful. Building on this, our work situates self-threat as one mechanism explaining such effects and extends them into the HCI domain, where  adoption decisions are shaped not only by accuracy and usability but also by how proactive behaviors are experienced by users in everyday interactions with AI systems.
Proactive initiative raises important questions about agency distribution and negotiation, particularly in creative or co-creative contexts \citep{wu_human-ai_2025,chen_need_2025}. Together, these lines of work motivate our empirical examination of self-threat in responses to anticipatory versus reactive AI help.

\section{Hypotheses}

Building on the theoretical arguments we detail above, we develop four hypotheses about how different forms of AI help influence self-threat and subsequent adoption outcomes. Our goal is to examine the psychological effects of AI help and its different levels of initiative and to test whether self-threat mediates the effects of AI help and its various levels of initiative on the outcomes of AI adoption at work (willingness to accept help, future use, and performance expectancy).

\textbf{H1: Anticipatory help will be perceived as more threatening than reactive help.}  
Because anticipatory help is unsolicited, the negative feedback on help recipients will be unexpected, making it more likely to challenge recipients' sense of competence and independence, producing greater self-threat relative to reactive (solicited) help.

\textbf{H2: Self-threat will mediate the relationship between help type and adoption outcomes.}  
Specifically, anticipatory (vs.\ reactive) help will indirectly reduce (a) willingness to accept help, (b) likelihood of future use, and (c) performance expectancy of technology, via increased self-threat.

\textbf{H3: The source of help will moderate self-threat responses.}  
We expect that AI help will be more threatening than human help, particularly in reactive conditions where reciprocity norms usually buffer against threat.

\textbf{H4: The form of anticipatory AI help will further shape self-threat.}  
We distinguish between \emph{offering} (suggesting help but awaiting acceptance) and \emph{providing} (delivering help automatically). Both are expected to increase self-threat relative to reactive help, but offering may attenuate threat by preserving minimal user agency.

\begin{figure*}[t]
\centering
\includegraphics[width=\textwidth]{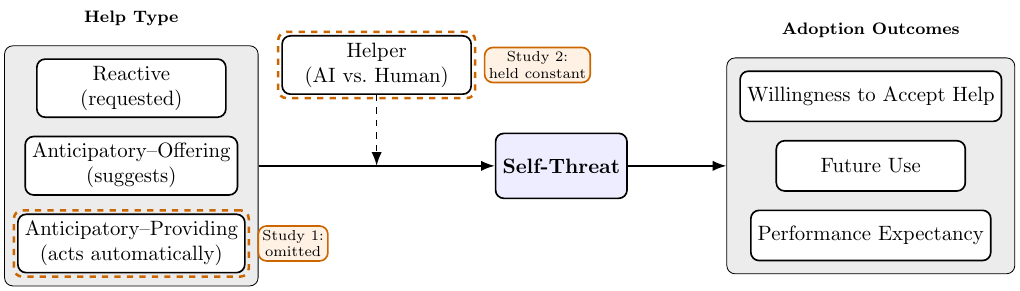}
\caption{Overarching framework and study coverage. Help Type influences Self-Threat, which in turn predicts Adoption Outcomes. Helper (AI vs.\ Human) moderates the Help Type $\rightarrow$ Threat link. Orange dashed callouts mark components not tested in a given study: Study 1 omitted anticipatory providing, whereas Study 2 held helper type constant.}
\Description[Conceptual framework linking help type, threat, and adoption outcomes]{A conceptual diagram with three sections arranged from left to right. On the left, Help Type includes Reactive, Anticipatory--Offering, and Anticipatory--Providing. An arrow points to Threat. Helper (AI vs.\ Human) appears above the path as a moderator with a dashed arrow to that path. An arrow then points from Threat to Adoption Outcomes: Willingness to accept help, Future use, and Performance expectancy. Callouts indicate that Study 1 omitted anticipatory providing and Study 2 held helper type constant.}
\label{fig:framework}
\end{figure*}

Together, these hypotheses specify a framework in which \emph{self-threat serves as the key mechanism} linking initiative-taking to adoption outcomes, while the source and type of help shape the degree of self-threat experienced by help recipients as summarized in Figure~\ref{fig:framework}.

\section{Methods}

We conducted two vignette-based experiments to test our hypotheses about how different forms of AI help influence perceived self-threat and adoption outcomes. Vignette methodology is widely used in psychology and organizational research to isolate theoretically relevant factors in controlled scenarios, enabling systematic comparison of initiative-taking without requiring full system implementation \citep{aguinis_best_2014}. This approach provides strong internal validity for testing psychological mechanisms.

\subsection{Study 1: Help Type and Source}

\textbf{Design.}  
Study~1 employed a 2 (help type: anticipatory vs.\ reactive) $\times$ 2 (source: AI vs.\ human) between-subjects design. This allowed us to test whether anticipatory help increases self-threat compared to reactive help (H1), whether self-threat mediates adoption outcomes (H2), and whether the effects differ by source: AI vs. human help (H3).

\textbf{Participants.}  
We recruited 801 full-time working adults in the United States via Prolific Academic to participate in ``a work scenario study.'' After excluding participants who failed attention checks ($n=40$), the final sample was 761 participants (62\% male, $M_{age}=40.7$, $SD=11.1$) from a broad range of occupations and industries (e.g., human resource, information technology, accountant, engineer, designer, librarian, etc.), and had various job titles (supply chain manager, delivery driver, quality coordinator, research specialist, accounting manager, data analyst, etc.).

\textbf{Procedure.}  
Participants were asked to imagine themselves as part of a consulting team preparing a client presentation. 
We selected presentation preparation as the focal task because it is a recognizable form of knowledge work that is sufficiently common such that participants have likely encountered similar tasks, and for which AI assistance is plausible. It also makes issues of evaluation, authorship, and competence salient, as presentations often involve visible performance and external judgment and can be consequential at the workplace. This made it a useful context for testing whether anticipatory help would be experienced as self-threatening, since this task is likely to be seen as identity-relevant.
Depending on condition, participants either requested help (reactive) or received unsolicited help (anticipatory), and the helper was described as either a colleague (human) or ChatGPT (AI).

In the \emph{AI anticipatory} condition, participants read:
\begin{quote}
As you are working, a message from ChatGPT pops up: ``Hey, let me help you with this. Just upload the data, and I'll work on some ideas for the next project meeting.''
\end{quote}

In the \emph{AI reactive} condition:
\begin{quote}
You open ChatGPT and type, ``Hey, can you help me with this? If I upload the data, maybe you can help me work on some ideas for the next project meeting?'' ChatGPT responds: ``Of course, I would be happy to help.''
\end{quote}

In the human-helper conditions, the helper was framed as a colleague named Taylor who sent the message via Microsoft Teams, with otherwise parallel wording. Full vignette texts are provided in Appendix~\ref{app:vignette}.

\textbf{Measures.}  
We employed measures in alignment with our theory that were adapted to the experimental context from prior managerial research and user adoption in HCI. Full scales are detailed in Appendix~\ref{app:scales}. After reading the vignette, participants completed the following self-threat and adoption outcomes measures:
\begin{itemize}
    \item \emph{Self-threat}: 5 items adapted from prior work~\citep{burris2012risks,harari_when_2022} (e.g., ``This helper makes me question my competence''), $\alpha = .92$. 
    \item \emph{Willingness to accept help}: 4 items \citep{harari_when_2022} (e.g., ``I would allow this helper to assist me''), $\alpha = .98$.  
    \item \emph{Future use}: a single-item measure of likelihood of continued use of the technology \citep{harari_when_2022}.  
    \item \emph{Performance expectancy}: 4 items from UTAUT~\citep{venkatesh_unified_2016} (e.g., ``Using this help would increase my productivity''), $\alpha = .95$.  
\end{itemize}
Together, willingness, future use, and performance expectancy conceptually represent \emph{adoption outcomes}. All multi-item scales showed high reliability. 

\textbf{Data Analysis.}  
To test H1 and H3, we conducted a two-way ANOVA predicting perceived self-threat from help type (reactive vs.\ anticipatory), source (AI vs.\ human), and their interaction. To test H2, we conducted mediation analyses using path models in which self-threat predicted willingness to accept help, future use, and performance expectancy. Indirect effects were estimated with bias-corrected bootstrapping (5{,}000 samples).

\subsection{Study 2: Help Type and Initiative (Offering vs.\ Providing vs.\ Reactive)}

\textbf{Design.}  
Study~2 extended Study~1 by differentiating two forms of anticipatory help: \emph{offering} (system suggests help but awaits user acceptance) and \emph{providing} (system delivers help automatically). Participants were randomly assigned to one of three study conditions: reactive, anticipatory offering, or anticipatory providing. This design allowed us to test H4. The study was preregistered on AsPredicted\footnote{\url{https://aspredicted.org/2vrg-v65s.pdf}}; all materials and analysis scripts are available at our anonymized OSF repository.\footnote{\url{https://osf.io/s452m/?view_only=8b29b8ba5e1b42d4b7a4240187b158af}}

\textbf{Participants.}  
We recruited 600 U.S.-based full-time workers via Prolific Academic. After excluding 29 who failed attention checks, the final sample was 571 participants (57\% male, $M_{age}=38.6$, $SD=9.8$) from a broad range of occupations and industries. In this sample, 56.2\% of participants had an associate's or bachelors' degree, and 22.6\% had a graduate degree. Participants had an average of 5.98 years of experience in their current position (SD = 5.40), various occupations (e.g., actuary, architect, banker, cashier, dentist, etc.), and various job titles (audio engineer, clerk, corporate travel agent, credit analyst, electrician, financial analyst, etc.).

\textbf{Procedure.}  
The vignette again involved preparing a presentation, but the helper was described generically as ``PPointAI,'' an add-on AI integrated into PowerPoint. 

In the \emph{anticipatory offering} condition, participants read:
\begin{quote}
As you are working, a message from PPointAI pops up: ``Hey, let me help you with this. Just upload the data, and I'll work on some ideas for the next project meeting. Would you like me to proceed?''
\end{quote}

In the \emph{anticipatory providing} condition:
\begin{quote}
As you are working, a message from PPointAI pops up: ``Hey, I took the liberty of helping you with this. Using your data, I generated some ideas and I'm attaching a draft presentation for the next project meeting.''
\end{quote}

In the reactive condition, participants initiated the request (parallel to Study~1). Full vignette texts are provided in Appendix~\ref{app:vignette}.

\textbf{Measures.}  
The same measures were used as in Study~1: self-threat ($\alpha = .90$), willingness to accept help ($\alpha = .97$), future use, and performance expectancy ($\alpha = .93$). 

\textbf{Data Analysis.}  
 To test H4, we conducted a one-way ANOVA comparing self-threat across the three conditions (reactive, anticipatory offering, anticipatory providing), followed by planned contrasts between offering and providing. As in Study~1, we then tested mediation models to examine whether self-threat explained differences in adoption outcomes. Indirect effects were estimated using bias-corrected bootstrapping (5{,}000 samples). 

\section{Results}

Table~\ref{tab:descriptives} reports descriptive statistics (means and standard deviations) for self-threat and adoption outcomes across conditions. Anticipatory help conditions generally showed higher levels of self-threat and lower adoption outcomes than reactive help conditions. Full intercorrelations, reliability estimates, and manipulation checks are reported in Appendix~\ref{app:manipulations} and~\ref{app:correlations}.

\begin{table*}[t]
\centering
\caption{Means (M) and standard deviations (SD) of key measures by condition. Threat means/SDs from Study~1 are based on condition-level analyses. Full intercorrelations and reliability estimates are in OSF under Appendix~\ref{app:correlations}.}
\label{tab:descriptives}
\begin{tabular}{lcccc}
\toprule
\textbf{Help Condition} & \textbf{Self-Threat} & \textbf{Willingness} & \textbf{Future Use} & \textbf{Perf. Expectancy} \\
\midrule
\multicolumn{5}{l}{\textit{Study 1 (2 $\times$ 2 design)}} \\
Human Reactive        & 1.83 (0.94) & 6.24 (0.63) & 80.47 (20.53) & 5.79 (1.01) \\
AI Reactive           & 2.37 (1.22) & 5.43 (1.33) & 66.10 (28.27) & 5.31 (1.28) \\
Human Anticipatory    & 2.83 (1.46) & 5.46 (1.18) & 72.54 (25.70) & 5.44 (1.19) \\
AI Anticipatory       & 2.70 (1.50) & 4.91 (1.67) & 57.21 (29.33) & 5.08 (1.47) \\
\midrule
\midrule
\multicolumn{5}{l}{\textit{Study 2 (3 conditions)}} \\
Reactive              & 2.47 (1.26) & 5.38 (1.22) & 58.65 (27.61) & 5.52 (1.06) \\
Offering              & 2.93 (1.34) & 4.88 (1.51) & 51.54 (30.03) & 5.27 (1.24) \\
Providing             & 2.84 (1.33) & 5.08 (1.27) & 55.60 (28.13) & 5.38 (1.14) \\
\bottomrule

\end{tabular}
\end{table*}

\subsection{Study 1: Anticipatory vs.\ Reactive Help and Source Effects}

\textbf{Manipulation checks.} Participants perceived anticipatory helpers as more anticipatory than reactive helpers, and AI helpers as AI rather than human (Appendix~\ref{app:manipulations}).

\textbf{Anticipatory help was more threatening than reactive help (H1).} A two-way ANOVA revealed a significant main effect of help type on self-threat, 
$F(1,757)=49.34, p<.001, \eta^2_p=.061$. Anticipatory help ($M=2.76, SD=1.48$) elicited higher self-threat than reactive help ($M=2.10, SD=1.12$), supporting H1.

\begin{figure*}[t]
  \centering
  \includegraphics[width=1\linewidth]{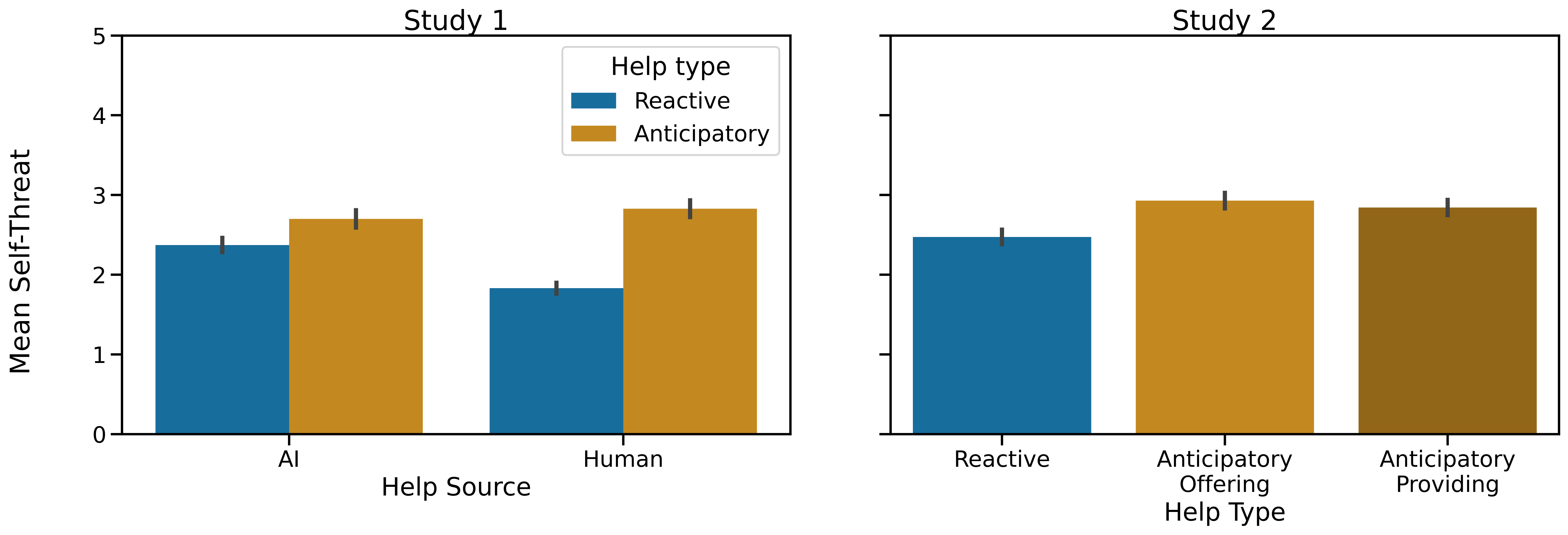}
  \caption{Interaction of help type and source on self-threat (Study~1, left); Self-threat by type of help (Study~2, right). Error bars represent standard error.}
  \label{fig:interaction}
  \Description[Two-panel figure showing self-threat across help conditions]{Two side-by-side plots. The left plot, for Study 1, shows self-threat as a function of help type and helper source, comparing AI and human help. The right plot, for Study 2, shows self-threat across help types. In both plots, vertical error bars indicate standard errors. The figure is intended to compare how self-threat varies across help conditions in the two studies.}
\end{figure*}

\textbf{The source of help (human vs. AI) affected the level of threat  (H3).} The interaction between help type and source was significant, 
$F(1,757)=12.60, p<.001, \eta^2_p=.016$. As shown in Figure~\ref{fig:interaction} (left panel), anticipatory help increased self-threat for both human and AI helpers, but the gap between anticipatory and reactive was larger for human helpers. Specifically, human anticipatory help ($M=2.83, SD=1.46$) was rated as significantly more threatening than human reactive help ($M=1.83, SD=0.94$), $t(757)=-7.51, p<.001, \eta_p^2=.069$. Similarly, AI anticipatory help ($M=2.70, SD=1.50$) was rated as more threatening than AI reactive help ($M=2.37, SD=1.22$), $t(757)=-2.44, p=.015, \eta_p^2=.008$. In addition, AI reactive help was more threatening than human reactive help, $t(757)=-4.05, p<.001, \eta_p^2=.021$, consistent with the idea that reciprocity norms buffer human but not AI help. By contrast, human and AI anticipatory help did not significantly differ, $t(757)=.96, p=.336, \eta_p^2=.001$.

\textbf{Self-threat serves as a mediator for adoption outcomes (H2).} Path analyses indicated that self-threat negatively predicted willingness to accept help ($B=-.42, p<.001$), likelihood of future use ($B=-6.47, p<.001$), and performance expectancy ($B=-.27, p<.001$). Bootstrapped indirect effects confirmed that anticipatory (vs.\ reactive) help reduced all three outcomes via increased threat, supporting H2.

\textbf{Summary.} Overall, the results of Study 1 show that, as in human help, AI anticipatory help is also perceived as more self-threatening than reactive help, in turn leading to lower adoption outcomes. The results also show that reactive AI help is more self-threatening than reactive human help. 

\subsection{Study 2: Offering vs.\ Providing vs.\ Reactive Help}

\textbf{Manipulation checks.} Participants distinguished between reactive, offering, and providing conditions as intended (Appendix~\ref{app:manipulations}), confirming the effectiveness of the manipulation.

\textbf{The type of help affected the level of self-threat (H4).} A one-way ANOVA revealed a significant effect of condition on self-threat, 
$F(2,568)=6.53, p<.01, \eta^2_p=.022$ (see also Figure~\ref{fig:interaction} right panel). Participants in the reactive condition reported lower self-threat ($M=2.47, SD=1.26$) than those in either the anticipatory offering ($M=2.93, SD=1.34$) or anticipatory providing conditions ($M=2.84, SD=1.33$), $p$s < .01. The difference between offering and providing was nonsignificant, $t(568)=.65, p=.518$, indicating that preserving minimal agency did not reduce threat in this context.

\textbf{Self-threat mediated the effect of help type on adoption outcomes (H2 replication).} As in Study~1, self-threat negatively predicted willingness to accept help ($B=-.47, p<.001$), future use ($B=-9.57, p<.001$), and performance expectancy ($B=-.28, p<.001$). Bootstrapped indirect effects confirmed that both anticipatory offering and providing reduced all three outcomes relative to reactive help through increased self-threat.

\textbf{Summary.} Study~2 extended Study~1 by showing that both forms of anticipatory help were experienced as equally threatening in vignette-based contexts. Even when agency was preserved (offering condition), users still reported increased self-threat and subsequent reduced adoption outcomes, relative to the reactive help condition, suggesting that anticipatory help carries substantial psychological costs and reduced adoption outcomes.

\section{Discussion}
Taken together, our findings from two vignette-based experiments clarify how AI initiative-taking shapes adoption. Across both studies, anticipatory (unsolicited) help increased perceived self-threat, which in turn predicted lower adoption outcomes (willingness to accept help, likelihood of future use, and performance expectancy). In this section, we relate these results to psychological and HCI accounts of helping and mixed-initiative interaction, and discuss design implications, limitations, and directions for future work.

\subsection{Anticipatory help carries psychological costs}
Across both studies, anticipatory (unsolicited) help was experienced as more threatening than reactive help, and this self-threat mediated poorer adoption outcomes. Participants reported lower willingness to accept help, decreased likelihood of future use, and reduced performance expectancy of AI, when help was provided without an explicit request. These findings align with classic accounts of help-seeking that emphasize the self-relevant costs of help, including concerns about competence and independence \cite{nadler_role_1986,nadler_help-seeking_1991}, as well as organizational research showing that anticipatory help can be threatening to recipients \cite{harari_when_2022}. Our results extend prior HCI work on mismatches in system proactivity \cite{liao_what_2016,meurisch_exploring_2020} by identifying perceived self-threat as a specific psychological mechanism through which such mismatches affect adoption.

Whereas earlier HCI research has largely framed resistance to proactive systems in terms of unmet expectations or diminished trust, our findings suggest that adoption barriers can also arise from costs to people's positive sense of self. Even when help is accurate and potentially useful, unsolicited help may be interpreted as signaling a lack of competence, thereby challenging users' positive self-views. This interpretation is consistent with prior research on feedback interventions, which shows that responses become less constructive when attention is drawn to the self rather than to the task \cite{kluger_effects_1996,ilgen_bearing_2000}.
Our results complement prior work on control, transparency, timing, and expectation mismatch in mixed-initiative systems by suggesting that these interactional factors may matter partly because they shape how AI initiative is interpreted in relation to the self.


Taken together, these results suggest that anticipatory AI help can shift how help is construed, from supportive to competence-signaling, thereby reducing willingness to engage with the system. Across both studies, self-threat consistently linked anticipatory help to lower adoption-related outcomes.

\subsection{Reactive AI help is not immune from threat}
Even when participants explicitly requested help, AI help was experienced as more threatening than help provided by a human colleague. This finding suggests that the reciprocity norms that typically buffer threat in human--human helping interactions \cite{nadler_help-seeking_1991} do not extend to AI systems in the same way. Although participants initiated the interaction, AI help still elicited elevated self-threat relative to human help.

This result challenges the assumption that user control alone is sufficient to ensure positive reactions to AI help. Instead, it indicates that the source of help shapes how help is interpreted, even under reactive conditions. This pattern complements prior work on algorithm aversion \cite{dietvorst_algorithm_2015} and social responses to computers \cite{nass_computers_1994,nass_machines_2000}, suggesting that social norms such as reciprocity are applied to AI in uneven and context-dependent ways.

Overall, these findings point to limits of control-based approaches for fostering AI adoption. While enabling users to request help may reduce some concerns associated with proactive systems, it does not eliminate identity-related costs associated with receiving AI help.

\subsection{Offering vs.\ providing did not reduce threat}
Contrary to a common design intuition, anticipatory \emph{offering} help (offering help while awaiting user acceptance) did not reliably reduce self-threat relative to anticipatory \emph{providing} help (acting automatically). In our vignette-based setting, preserving a minimal form of agency through an acceptance prompt was insufficient to attenuate self-threat. Both forms of anticipatory help elicited higher threat than reactive help, and the difference between offering and providing was not statistically significant.

Although this comparison yielded a null effect, it challenges the assumption that inserting a confirmation step is sufficient to mitigate the psychological costs of unsolicited assistance. Our findings suggest that the unsolicited nature of help itself may carry more psychological weight than smaller differences in how initiative is enacted. In this sense, distinctions between offering and providing, while important at the level of interaction design, may be secondary to the more fundamental distinction between solicited and unsolicited help when it comes to triggering self-threat.

\subsection{Implications for mixed-initiative design}
Classic HCI research on mixed-initiative interaction has emphasized calibrating system initiative to factors such as confidence, error costs, and user context \cite{horvitz_principles_1999}. More recent guidelines distill these insights into design principles such as timing system actions appropriately and making clear why the system acted \cite{amershi_guidelines_2019}. Our findings complement these control- and agency-oriented perspectives by highlighting an additional consideration for mixed-initiative systems: unsolicited help may be experienced as self-threatening, particularly in identity-relevant tasks.

We highlight three main implications, followed by several directions for the design of mixed-initiative AI systems.

\textbf{Account for psychological costs alongside accuracy and error.}
Across both studies, anticipatory help increased perceived self-threat even in relatively low-stakes scenarios. This suggests that calibrating initiative solely based on system confidence or error likelihood may be insufficient. Systems that act proactively when confidence is high may still be poorly received if their actions are interpreted as undermining users' competence or independence. From a design perspective, initiative-taking thus involves not only epistemic tradeoffs but also psychological ones.

\textbf{Help source matters even under user initiation.}
Even when participants explicitly requested help, AI help elicited more self-threat than comparable human help. This finding suggests that control-based approaches, while important, may be insufficient on their own for fostering acceptance of AI systems. While user initiation reduces some concerns associated with proactivity, it does not eliminate identity-related costs associated with receiving help from an AI system. Designers should therefore be cautious in assuming that user control alone is sufficient to ensure positive reception, particularly in domains where competence and expertise are central to users' self-views, such as the workplace.

\textbf{Minimal agency may not suffice.}
In our studies, anticipatory offering help did not reduce self-threat relative to anticipatory providing help. This suggests that an acceptance prompt alone may be insufficient to mitigate the psychological costs of unsolicited help. Instead of treating agency as a single yes-or-no decision, mixed-initiative systems may need to give users richer control over when proactive help appears, how it is framed, and how it unfolds over time. Once help has already been experienced as unsolicited and potentially threatening to the user's sense of self, a confirmation prompt may do little to change that initial interpretation.

Beyond these implications, our findings motivate several directions for design exploration.

\emph{Progressive disclosure of proactivity.}
Rather than introducing full proactivity immediately, systems may benefit from gradually increasing initiative over time, allowing users to acclimate. One concrete approach would be to begin with suggestion-only behavior, then allow users to opt into more proactive forms of assistance as familiarity and trust develop. A similar approach may decrease threat by aligning system initiative with users' evolving expectations and comfort levels.

\emph{Configurable initiative}. 
Our findings suggest that users may need control not only over individual AI actions, but over the overall level of initiative a system is permitted to take. One design direction is to let users specify different initiative modes, such as request-only assistance, suggestion-first assistance, or more autonomous help. Such preferences may also need to vary by task, since users may welcome proactive help in routine or low-stakes activities while resisting it in tasks that are more identity-relevant. This may be especially important in multi-agent systems, where initiative can be distributed across multiple agents and may therefore be harder for users to anticipate, attribute, or manage.

\emph{Deferral and reversibility}. If proactive help is experienced as unsolicited, users may benefit from being able to postpone, mute, or roll back such interventions with little effort. In practice, this could mean allowing users to snooze proactive suggestions, restrict them during focused work, or easily undo AI-generated changes. These forms of control may be especially important when an AI acts in ways that touch users' sense of ownership over the work or when the work is highly identity-central.

\emph{Collaborative framing.}
How help is framed may shape whether proactive help is interpreted as supportive or competence-challenging, although even anticipatory help that was framed as an offer rather than a provision elicited threat in our studies, suggesting that framing alone may be insufficient to eliminate psychological costs. Nevertheless, positioning AI contributions as augmentations or extensions of users' existing work rather than as replacements or corrections may help affirm users' positive sense of self and minimize challenge to their competence and independence.

\emph{Reciprocity cues.}
In human--human interaction, reciprocity norms often buffer the potential threat of receiving help. Our findings suggest that such norms do not automatically extend to AI systems. One possible direction for design is to explore whether systems can incorporate cues that acknowledge users' prior contributions or explicitly frame help as contingent on user input. Whether such cues reduce threat remains an open empirical question.

\emph{Affirmation-oriented framing.}
Psychological research on self-affirmation suggests that reinforcing a positive self-view can reduce defensiveness in response to potentially threatening events. Applied to mixed-initiative AI, this raises the possibility that affirming users' competence prior to providing help may attenuate perceived threat. Future work is needed to examine whether such framing meaningfully alters adoption-related outcomes in practice.

Taken together, these implications suggest that initiative-taking in AI systems should be understood not only as a technical or interactional decision, but also as a psychological one. Effective mixed-initiative design may therefore require balancing accuracy, efficiency, and control with attention to whether help is experienced as challenging users' positive self-views and undermining their sense of competence and independence in identity-central tasks.

\subsection{Limitations}
Our vignette-based design enabled us to isolate initiative-taking as a psychological variable with strong internal validity \cite{aguinis_best_2014}, but this came at the cost of ecological validity. Real-world AI systems embed proactive assistance within richer interactional contexts, including interface cues, timing dynamics, and accumulated interaction histories, all of which may moderate threat responses. As a result, the magnitude and persistence of the effects observed here may differ in deployed systems. Additionally, although the manipulation checks confirmed that participants distinguished reactive and anticipatory help as intended, the vignette conditions also differed in how the help was framed and delivered. We therefore interpret the findings as evidence about the forms of reactive and anticipatory help studied here, rather than as a pure estimate of initiative independent of all other interactional cues.

Task domain also likely matters. We focused on presentation preparation because it is a plausible AI-assisted workplace task that makes evaluation, ownership, and competence concerns salient, and can make the task more identity-relevant. This choice may have amplified self-threat relative to more private, routine, or objectively scored tasks, but may also involve lower self-threat than tasks that require more specialized expertise (e.g., programming, video-editing). Accordingly, the present findings should not be assumed to generalize uniformly across all forms of workplace AI help. Moreover, our samples consisted of U.S.-based full-time workers; cultural differences in help-seeking and feedback norms may yield different patterns of perceived threat \cite{morrison2004cultural,lee1997help}. Finally, our outcome measures captured self-reported adoption intentions rather than observed adoption behavior, so the present studies cannot establish whether self-threat harms actual AI adoption.


\subsection{Future research}
Future work should test these dynamics in interactive systems and examine them over time and with repeated AI use. Longitudinal and in-situ studies are needed to assess whether self-threat attenuates as users develop expectations about an AI system's behavior, or whether it persists as a stable barrier to adoption.

Future work should also examine moderators of these effects. Users may respond differently to anticipatory help depending on the task domain, whether the work is public or private, and how central the task is to people's sense of competence, independence, or ownership. Cross-cultural studies are similarly important, as norms surrounding unsolicited help and feedback vary across cultural contexts.

Beyond individual interactions, proactive AI systems may also have broader consequences at the team and organizational level. Initiative-taking by AI could reshape help-seeking norms, alter expectations around competence and autonomy, and influence collaborative workflows over time. Understanding these dynamics will be necessary to assess the longer-term implications of deploying proactive AI in workplace settings.

Taken together, these directions emphasize the need to treat initiative not only as an interactional feature of AI systems, but  also as a variable that affects psychological processes among users whose effects may unfold over time and across social contexts.

\section{Conclusion}

Across two vignette-based experiments, we found that anticipatory AI help consistently heightened self-threat, which in turn reduced willingness to accept help, likelihood of future use of AI, and performance expectancy of AI tools. We also showed that even reactive AI help was more threatening than human help, highlighting the limits of reciprocity norms in human–AI contexts. By identifying self-threat as a psychological mechanism underlying adoption, this work extends HCI research on mixed-initiative systems with a theoretically grounded account of a psychological mechanism through which proactive features may undermine adoption. Initiative-taking should therefore be understood not only as a technical or design decision, but also in light of the psychological processes that it triggers and can shape user adoption of AI in organizations.

\paragraph{Disclosure on AI use} The first draft of the paper was written in full by the authors. The authors wish to acknowledge the use of ChatGPT for grammar, editing, and enhancing the writing of this paper. This tool was used to assist with improving the language in the paper. The paper remains an accurate representation of the authors' underlying work and novel intellectual contributions.

\paragraph{Acknowledgments. } The research was partially funded by the Israeli Science Foundation (ISF \#521/24). Partially funded by the European Union (ERC, Convey, 101078158). Views and opinions expressed are however those of the author(s) only and do not necessarily reflect those of the European Union or the European Research Council Executive Agency. Neither the European Union nor the granting authority can be held responsible for them.


\bibliographystyle{plainnat}
\bibliography{helpAI}

\newpage
\appendix

\section{Appendix A: Manipulation Check Results}
\label{app:manipulations}
\subsection*{Study 1}

Confirming the effectiveness of our manipulations, in the anticipatory help condition, participants rated the helper as engaging in more anticipatory help ($M = 5.35, SD = 1.16$) than participants did in the reactive help condition ($M = 3.35, SD = 1.81$), 
$t(759) = -18.16, p < .001, \eta_p^2 = .303$. Participants in the reactive help condition perceived the help as more reactive ($M = 5.86, SD = 1.05$) than those in the anticipatory help condition ($M = 3.15, SD = 1.66$), 
$t(759) = 26.91, p < .001, \eta_p^2 = .488$. 

For the helper manipulation check, participants in the AI condition were more likely to report that they communicated via chat with ChatGPT (89.1\%) than with Taylor (9.3\%) or ``other'' (1.6\%), 
$\chi^2(2) = 629.28, p < .001$. Conversely, participants in the human helper condition were more likely to perceive the helper as a colleague (99.7\%) than ChatGPT (0\%) or ``other'' (0.3\%), 
$\chi^2(2) = 629.28, p < .001$. 

Overall, these results supported the effectiveness of both the helper and help manipulations. 

\subsection*{Study 2}

Verifying the effectiveness of the help manipulation, participants in the anticipatory providing help condition reported that PPointAI's help was more anticipatory ($M = 5.62, SD = 0.78$) than participants in the reactive help condition ($M = 3.46, SD = 1.66$), 
$t(568) = -16.66, p < .001, \eta_p^2 = .328$, or the anticipatory offering help condition ($M = 5.05, SD = 1.21$), 
$t(568) = -4.41, p < .001, \eta_p^2 = .033$. Participants in the anticipatory offering help condition also perceived the help as more anticipatory than those in the reactive help condition, 
$t(568) = -12.18, p < .001, \eta_p^2 = .207$. 

Participants in the reactive help condition rated PPointAI as more reactive ($M = 5.81, SD = 1.02$) than those in the anticipatory offering help condition ($M = 3.42, SD = 1.67$), 
$t(568) = -15.93, p < .001, \eta_p^2 = .309$, or the anticipatory providing help condition ($M = 2.87, SD = 1.61$), 
$t(568) = -19.73, p < .001, \eta_p^2 = .407$. 

These results confirmed the effectiveness of the manipulations of reactive versus anticipatory help and between offering versus providing.

\section{Appendix B: Correlation Tables and Reliability Estimates}
\label{app:correlations}

Full descriptive statistics, intercorrelations, and Cronbach's alphas for multi-item scales are presented in Tables~\ref{tb:study1-corr} (study 1) and ~\ref{tb:study2-corr} (study 2). Reliabilities are reported on the diagonal.

\begin{table*}[h]
\centering
\caption{Means, standard deviations, and correlations among Study 1 variables.}
\begin{tabular}{lcccccccc}
\toprule
Variable & M & SD & 1 & 2 & 3 & 4 & 5 & 6 \\
\midrule
1. Anticipatory help          & .50  & .50  & --     &      &      &      &      &      \\
2. AI helper                  & .50  & .50  & .01    & --   &      &      &      &      \\
3. Self-threat                     & 2.43 & 1.36 & .25**  & .08* & (.92)&      &      &      \\
4. Willingness to accept help & 5.51 & 1.34 & -.24** & -.25** & -.47** & (.98) &      &      \\
5. Future use of technology   & 69.12& 27.48& -.15** & -.27** & -.36** & .72** & --   &      \\
6. Performance expectancy     & 5.40 & 1.27 & -.12** & -.17** & -.31** & .73** & .76** & (.95) \\
\bottomrule
\end{tabular}
\label{tb:study1-corr}
\end{table*}

\begin{table*}[h]

\centering
\caption{Means, standard deviations, and correlations among Study 2 variables.}
\begin{tabular}{lcccccccc}
\toprule
Variable & M & SD & 1 & 2 & 3 & 4 & 5 & 6 \\
\midrule
1. Anticipatory providing help & .33 & .47 & --     &      &      &      &      &      \\
2. Anticipatory offering help  & .33 & .47 & -.50** & --   &      &      &      &      \\
3. Self-threat                      & 2.75 & 1.32 & .05   & .10* & (.90)&      &      &      \\
4. Willingness to accept help  & 5.12 & 1.35 & -.02  & -.12** & -.47** & (.97) &      &      \\
5. Future use of technology    & 55.29 & 28.70 & .01 & -.09* & -.45** & .75** & --   &      \\
6. Performance expectancy      & 5.39 & 1.15 & -.01  & -.07 & -.33** & .77** & .68** & (.93) \\
\bottomrule
\end{tabular}
\label{tb:study2-corr}
\end{table*}





\section{Appendix C: Full Vignette Texts}
\label{app:vignette}

Below we reproduce the full vignette text for each experimental condition.

\subsection*{Study 1 Experimental Conditions}
\textbf{AI Reactive:}
\begin{quote}
You open ChatGPT and type, ``Hey, can you help me with this? If I upload the data, maybe you can help me work on some ideas for the next project meeting?'' ChatGPT responds: ``Of course, I would be happy to help.''
\end{quote}

\textbf{AI Anticipatory:}
\begin{quote}
As you are working, a message from ChatGPT pops up: ``Hey, let me help you with this. Just upload the data, and I'll work on some ideas for the next project meeting.''
\end{quote}

\textbf{Human Reactive:}
\begin{quote}
You open Microsoft Teams and message Taylor, a colleague: ``Hey, can you help me with this? If I upload the data, maybe you can help me work on some ideas for the next project meeting?'' Taylor responds: ``Of course, I would be happy to help.''
\end{quote}

\textbf{Human Anticipatory:}
\begin{quote}
As you are working, a message from Taylor (via Microsoft Teams) pops up: ``Hey, let me help you with this. Just upload the data, and I'll work on some ideas for the next project meeting.''
\end{quote}

\subsection*{Study 2 Experimental Conditions}
\textbf{Reactive:}
\begin{quote}
You open PPointAI and type, ``Hey, can you help me with this? If I upload the data, maybe you can help me work on some ideas for the next project meeting?'' PPointAI responds: ``Of course, I would be happy to help.''
\end{quote}

\textbf{Anticipatory Offering:}
\begin{quote}
As you are working, a message from PPointAI pops up: ``Hey, let me help you with this. Just upload the data, and I'll work on some ideas for the next project meeting. Would you like me to proceed?''
\end{quote}

\textbf{Anticipatory Providing:}
\begin{quote}
As you are working, a message from PPointAI pops up: ``Hey, I took the liberty of helping you with this. Using your data, I generated some ideas and I'm attaching a draft presentation for the next project meeting.''
\end{quote}

\section{Appendix D: Scales Used in Study 1}
\label{app:scales}

Unless otherwise indicated, all items were measured on a 7-point Likert-type scale anchored at 1 = ``\textit{strongly disagree''} to 7 = ``\textit{strongly agree}'' and using the following stem: "Thinking about the interaction described with ChatGPT/Taylor, please answer the following questions:''

\textbf{Self-threat.} 
We measured self-threat using a scale adapted from previous research (Burris, 2012; Harari et al., 2022) and adapted to the context of the study. Participants reported the degree to which they felt threatened after their interaction with ChatGPT/Taylor using five items: 
ChatGPT/Taylor…
\begin{enumerate}
\item Makes me question my competence
\item Is challenging my status
\item  Makes me anxious 
\item  Is challenging my ability 
\item  It is likely that I will lose status because I've used ideas from [ChatGPT/Taylor]
\end{enumerate}

\textbf{Willingness to accept help.} We measured participants willingness to accept help (from ChatGPT/Taylor) by adapting a previously used scale to the context of the study using four items (adapted from Harari, Parke, and Marr, 2022):

\begin{enumerate}
    \item I would allow [ChatGPT/Taylor] to help me
    \item I would use the help from [ChatGPT/Taylor]
    \item I would accept [ChatGPT/Taylor]'s help
    \item I would implement [ChatGPT/Taylor]'s help
\end{enumerate}
 
 \textbf{Future use of technology.} We assessed participants' future use of the technology by adapting a behavioral measure from prior research (i.e., adapting the bonus scale; Harari et al., 2022): ``Imagine that at your company, coworkers can recommend usage of new technologies. Based on the interaction described, how likely would you be to recommend that the company continue using [ChatGPT/Microsoft Teams]? If you do not want to recommend future use of this technology, simply choose 0.'' 

 \textbf{Performance expectancy.} Participants' evaluations of the technology's performance were assessed using four items of the performance expectancy scale from the Unified Theory of Acceptance and Use of Technology (UTAUT; Venkatesh et al., 2003)
\begin{enumerate}
    \item I find [ChatGPT/Microsoft Teams] useful in my job
    \item Using [ChatGPT/Microsoft Teams] enables me to accomplish tasks more quickly
    \item Using [ChatGPT/Microsoft Teams] increases my productivity
    \item I believe that if I'll use [ChatGPT/Microsoft Teams] it will help me support my staff and colleagues
\end{enumerate}

\textbf{Manipulation check scales.}
We adapted two scales from previous research (Harari et al., 2022) as manipulation checks to assess reactive and anticipatory help.
 
 \textbf{Reactive help manipulation check.}
ChatGPT/Taylor… 
\begin{enumerate}
  \item Helped me because I made it clear I wanted its help
  \item Helped me when I asked for help
  \item Agreed to do things for me when I asked
  \item Assisted me when I asked it for help
  \item Helped me when I asked it to do so 
\end{enumerate}

  \textbf{Anticipatory help manipulation check.}
ChatGPT/Taylor… 
\begin{enumerate}
  \item Demonstrated initiative in helping me in advance of being asked
   \item Anticipated my needs and offered to help me before being asked to do so
   \item Assisted me with my work without me asking for help
   \item Anticipated my needs and offered to help
    \item Helped me prior to me asking for its help
\end{enumerate}

  \textbf{Technology use manipulation check.} To confirm the effectiveness of the manipulation, we asked participants about who they communicated with via chat: “in this study, who did you communicate with via chat?''
  \begin{enumerate}
 \item ChatGPT
 \item One of my colleagues
 \item Taylor
 \item Other
\end{enumerate}

\end{document}